\documentclass{pjab}%           <- for two column
\usepackage[T1]{fontenc}
\usepackage{lmodern}
\usepackage{textcomp}
\usepackage{amsmath, amssymb}
\usepackage{graphicx}
\usepackage{array}
\usepackage[superscript]{cite}
\usepackage{multirow, threeparttable}
\usepackage{ulem}

\newcommand\farcs{\mbox{$.\!\!^{\prime\prime}$}}% 
\newcommand\arcsec{\mbox{$^{\prime\prime}$}}% 
\newcommand\sun{\odot}% 

\newcommand\aj{AJ}%        % Astronomical Journal 
\newcommand\araa{ARA\&A}%  % Annual Review of Astron and Astrophys 
\newcommand\apj{ApJ}%    % Astrophysical Journal 
\newcommand\apjl{ApJL}     % Astrophysical Journal, Letters 
\newcommand\apjs{ApJS}%    % Astrophysical Journal, Supplement 
\newcommand\aap{A\&A}%     % Astronomy and Astrophysics 
\newcommand\aapr{A\&A~Rv}%  % Astronomy and Astrophysics Reviews 
\newcommand\mnras{MNRAS}% % Monthly Notices of the RAS
\newcommand\nat{Nature}% % Nature
\newcommand\pasj{PASJ}% % Publications of the ASJ

\begin{document}
%\Vol{98}
%\No{3}
\pjabcategory{Original Article}
%\pjabcategory{Original Article}
\title[The origin of a star around SMBH]
%      {S0-6/S10: Orbit, Metallicity, and Age of an Orbiting Star Around the Galactic Supermassive Black Hole}
      {Origin of an Orbiting Star Around the Galactic Supermassive Black Hole}
%\subtitle{subtitle}

\authorlist{%
\Cauthorentry{Shogo Nishiyama}{labelA}% <- correspondending author
\authorentry{Tomohiro Kara}{labelA}
\authorentry{Brian Thorsbro}{labelB}
\authorentry{Hiromi Saida}{labelC}
\authorentry{Yohsuke Takamori}{labelD}
\authorentry{Masaaki Takahashi}{labelE}
\authorentry{Takayuki Ohgami}{labelF}
\authorentry{Kohei Ichikawa}{labelG, labelH}
\authorentry{Rainer Sch\"{o}del}{labelI}
}

%\affiliation[labelA]{Miyagi University of Education, 149 Aramaki-Aza-Aoba, Aoba-ku, Sendai, Miyagi 980-0845, Japan}
\affiliate[labelA]{Miyagi University of Education, Sendai, Miyagi, Japan}
\affiliate[labelB]{Observatoire de la C\^{o}te d'Azur, CNRS UMR 7293, BP4229, Laboratoire Lagrange, F-06304 Nice Cedex 4, France}
\affiliate[labelC]{Daido University, Naogya, Aichi, Japan}
\affiliate[labelD]{National Institute of Technology, Gobo, Wakayama, Japan}
\affiliate[labelE]{Aichi University of Education, Kariya, Aichi, Japan}
\affiliate[labelF]{National Astronomical Observatory of Japan, Tokyo, Japan}
\affiliate[labelG]{Tohoku University, Sendai, Miyagi, Japan}
\affiliate[labelH]{Waseda University, Tokyo, Japan}
\affiliate[labelI]{Instituto de Astrof\'{i}sica de Andaluc\'{i}a (IAA)-CSIC, Granada, Spain}

\Correspondence{Shogo Nishiyama, Miyagi University of Education, 149 Aramaki-Aza-Aoba, Aoba-ku, Sendai, Miyagi 980-0845, Japan, (shogo-n@staff.miyakyo-u.ac.jp)}
%\Abbreviations{PJA: Proceedings of the Japan Academy; RNA: Ribonucleic acid}

\abstract{
The tremendous tidal force that is linked to the supermassive black hole (SMBH) at the center of our galaxy is expected to strongly subdue star formation in its vicinity. Stars within $1\arcsec$ from the SMBH thus likely formed further from the SMBH and migrated to their current positions. In this study, spectroscopic observations of the star S0-6/S10, one of the closest (projected distance from the SMBH of $\approx 0\farcs3$) late-type stars were conducted. Using metal absorption lines in the spectra of S0-6, the radial velocity of S0-6 from 2014 to 2021 was measured, and a marginal acceleration was detected, which indicated that S0-6 is close to the SMBH. The S0-6 spectra were employed to determine its stellar parameters including temperature, chemical abundances ([M/H], [Fe/H], [$\alpha$/Fe], [Ca/Fe], [Mg/Fe], [Ti/Fe]), and age. As suggested by the results of this study, S0-6 is very old ($\gtrsim 10$\,Gyr) and has an origin different from that of stars born in the central pc region.
}
%%\ed-mja{Edited by xxx \textsc{yyy}, ...}
%%\received{2021}{7}{28}
%%\accepted{2021}{8}{31}
%%\doi{10.2183}{96.014}{
\maketitle

%%%%%%%%%%%%%%%%%%%%%%%%%%%%%%%%%%%%%%%%%%%%%%%%%%%%%%%
%%%%%%%%%%%%%%%%%%%%%%%%%%%%%%%%%%%%%%%%%%%%%%%%%%%%%%%
\section{Introduction}\label{sec:intro}

Stars within approximately 1\arcsec (
$\approx 0.04$ pc
 at a distance of 8\,kpc from the Earth) from the supermassive black hole (SMBH) at the center of our galaxy are referred to as ``S-stars''\footnote{There is no common definition for S-stars. Here, S-stars are defined as those within $1\arcsec$ from Sgr\,A*, which include both early- and late-type stars.}. S-stars are a collection of both early-type and late-type stars, and the early-type are B0--B3 main sequence stars with masses of $8$--$14\,M_{\sun}$ \cite{2003ApJ...586L.127G,2017ApJ...847..120H}. The late-type stars are G--M red giants with estimated initial masses of $0.5$--$2\,M_{\sun}$ \cite{2019ApJ...872L..15H}.

\begin{figure*}[h]
\begin{center}
   \includegraphics[width=9cm]{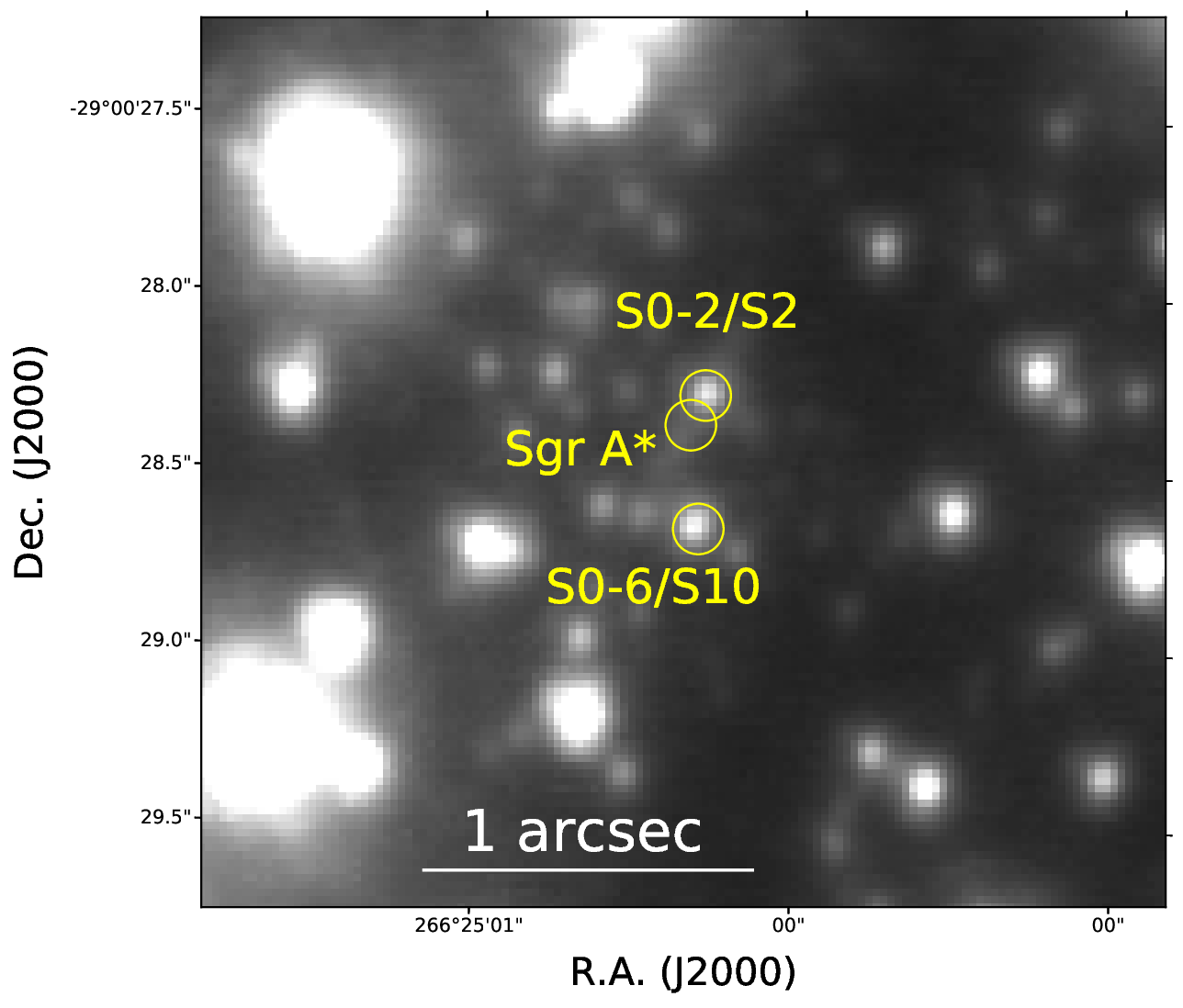}
    \caption{$K'$-band image of the galactic center region
            taken with Subaru/IRCS/AO188 in 2017.
            North is up, and east is to the left in the image.
            The positions of S0-2/S2, Sgr\,A*,
            and our target S0-6/S10, are shown by yellow circles.
            \label{fig:KsImS0-6}}
            \end{center}
            \vspace{-0.6cm}
\end{figure*}

\begin{figure*}[t]
\vspace{-1.0cm}
\begin{center}
    \includegraphics[angle=90,width=16cm]{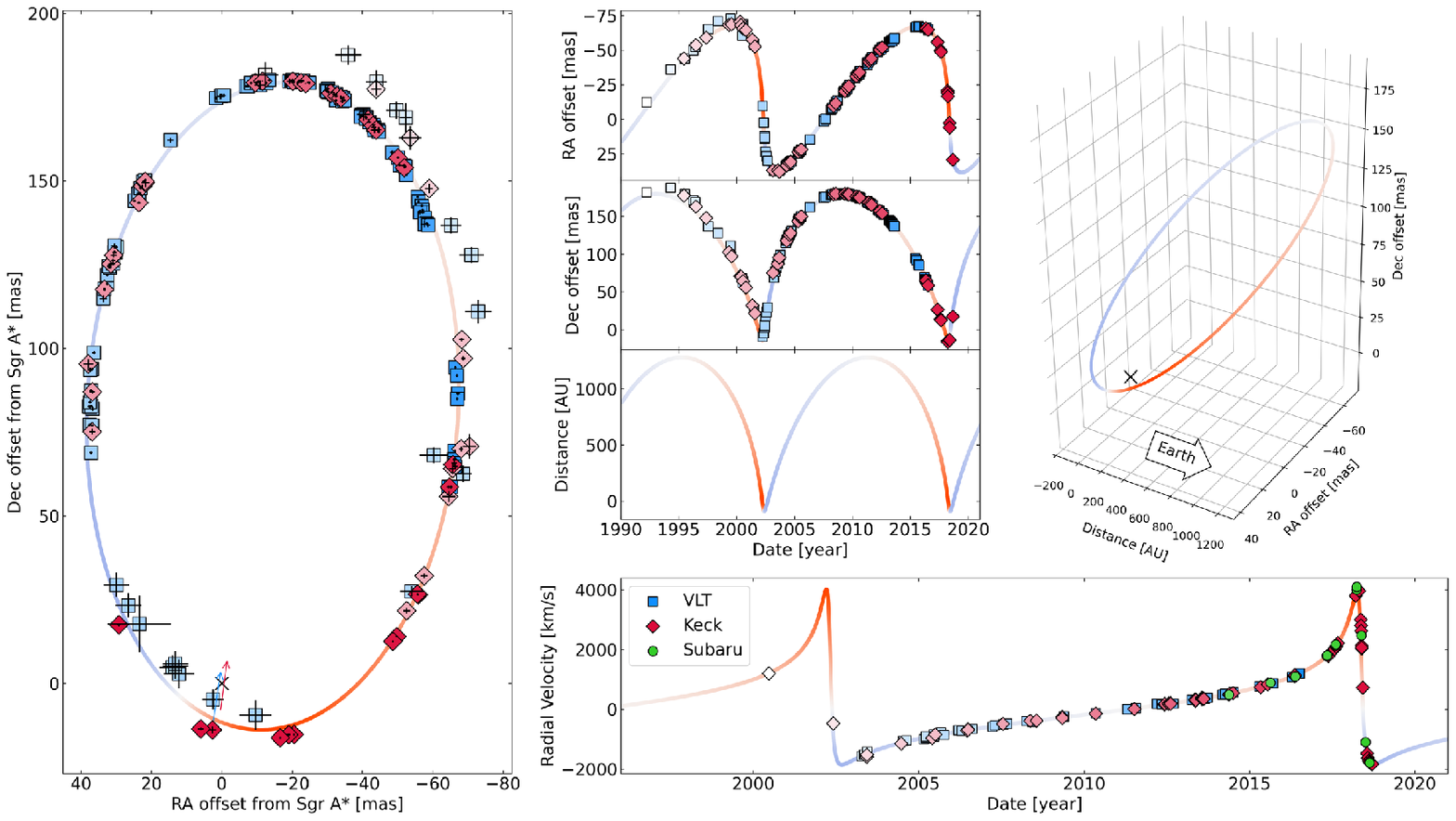}
    \vspace{-1.2cm}
    \caption{Observational results and derived orbit of S0-2/S2.
        The blue squares, red diamonds, and green circles respectively denote the results
        of the VLT, Keck/Gemini, and Subaru observations.
        The darkness of the colors indicates the epoch of observations,
        with the most recent ones represented by the darkest points.
        The results of an orbital fitting are shown by blue and red lines,
        which indicate the blue- and red-shifted parts of the orbit, respectively.
        (Left) Orbital motion of S0-2 on the sky.
        The positions are relative to Sgr\,A*.
        S0-2/S2 revolves around Sgr\,A* in a clockwise direction in this plot.
        (top middle) From top to bottom, 
        offset in right ascension (RA), offset in declination (Dec),
        and distance from Sgr\,A*, respectively, as function of time.
        (top right) Three-dimensional visualization of the orbit of S0-2/S2.
        The direction to the Earth is indicated by the white arrow,
        and the position of Sgr\,A* is depicted by the black cross.
        (bottom right) Radial velocity of S0-2 as function of time.
            \label{fig:OrbitS0-2}}
            \end{center}
\end{figure*}

S-stars provide a unique test bed for probing the strong gravity around SMBHs. Owing to the large mass ratio between S-stars and the SMBH, named Sagittarius A* (Sgr\,A*; $\approx 4 \times 10^6 M_{\sun}$\cite{2002Natur.419..694S}), the stars can be considered as test particles that move in a static potential. A particularly interesting target is the star named S0-2 or S2 (Fig.\,\ref{fig:KsImS0-6}),
whose orbital period is $\approx 16$\,years \cite{2002Natur.419..694S, 2003ApJ...586L.127G}. We have been carrying out spectroscopic monitoring observations since 2014 with the aim to precisely monitor the radial velocity (RV) of S0-2/S2 around the pericenter passage in 2018 \cite{2018PASJ...70...74N}. We were able to spatially resolve S0-2/S2 and obtain high-resolution near-infrared (NIR) spectra with the use of the infrared camera and spectrograph (IRCS) \cite{2000SPIE.4008.1056K} on the Subaru telescope \cite{2004PASJ...56..381I}, in combination with the Subaru adaptive optics (AO) \cite{2010SPIE.7736E..0NH} and laser guide star (LGS) systems \cite{2012SPIE.8447E..1FM}. The observational results and the obtained orbit of S0-2/S2 are shown in Fig.\,\ref{fig:OrbitS0-2}.

The observations and analysis provide not only the orbital parameters for S0-2/S2 but also the physical parameters for the SMBH. In our study, the mass and distance of the SMBH were determined to be $(4.23 \pm 0.07) \times 10^6\,M_{\sun}$ and $8.10 \pm 0.07$\,kpc, respectively (Saida et al.\cite{2019PASJ...71..126S}; see also \cite{2019A&A...625L..10G, 2019Sci...365..664D}).
It should be noted that by observing stars that orbit the SMBH, we can determine the mass of the SMBH with an uncertainty of only a few percent, which is much smaller than those from observations of the shadow of the SMBH \cite{2022ApJ...930L..12E}.

%%%%%%%%%%%%%%%%%%%%%%%%%%%%%%%%%%%%%%%%%%%%%%%%%%%%%%%
%%%%%%%%%%%%%%%%%%%%%%%%%%%%%%%%%%%%%%%%%%%%%%%%%%%%%%%

Because of the tremendous tidal force associated with the SMBH, it is highly unlikely that the S-stars formed at their present position \cite{2016LNP...905..205M}.
Therefore, the stars are generally assumed to form at larger distances from Sgr\,A* and were brought to their current location via dynamical processes. Considering that the late-type S-stars are likely to be old, $\sim 3$--$10$\,Gyr \cite{2019ApJ...872L..15H}, they are expected to be dynamically relaxed. In other words, identifying where they formed using only kinematic information is impossible. In this case, stellar properties including age and chemical abundances are crucial in order to understand their origin.

The SMBH and S-stars are surrounded by a dense stellar system, the nuclear star cluster (NSC).
This is a massive ($\sim 2.5 \times 10^7 M_{\sun}$) cluster with a half-light radius of $\sim 5$\,pc
\cite{2014CQGra..31x4007S, 2020A&ARv..28....4N}.
Recent observations revealed that, besides a major stellar population with a mean metallicity at least as high as twice the solar, there exists a subpopulation of stars that are characterized by sub-solar metallicity \cite{2015ApJ...809..143D, 2017MNRAS.464..194F}.
The subsolar population may be a relic of an old star cluster migrated to the center of our galaxy \cite{2020MNRAS.494..396F, 2020ApJ...901L..28D}. In this paper, ``NSC stars'' is defined as stars within $\sim 5$\,pc from Sgr\,A* except S-stars. Habibi et al.\cite{2019ApJ...872L..15H} obtained spectra of 21 late-type S-stars within $0.5\,$pc from Sgr\,A*, and estimated their $T_{\mathrm{eff}}$, spectral type, age, and initial mass. However, owing to the medium resolution ($\sim 1500$) of the spectra, they could not determine the chemical abundance of the late-type S-stars. Without metallicity information, stellar ages are impossible to determine precisely.
In this study, we obtain high-resolution spectra of one of the late-type S-star S0-6 and determine its chemical abundances and age to investigate its origin.
S0-6 is located at $\approx 0\farcs3$ from Sgr\,A*
(Fig.\,\ref{fig:KsImS0-6}),
which corresponds to a transverse distance of 
$\approx 0.012\,$pc or $2400\,$AU.
S0-6 is of great interest because it appears to be one of the rare red giants 
that are very close to Sgr\,A*.

%%%%%%%%%%%%%%%%%%%%%%%%%%%%%%%%%%%%%%%%%%%%%%%%%%%%%%%
%%%%%%%%%%%%%%%%%%%%%%%%%%%%%%%%%%%%%%%%%%%%%%%%%%%%%%%
\section{Observations and data reduction} \label{sec:obs}

In this study, NIR spectroscopic observations were conducted using the Subaru telescope and IRCS in the echelle mode (spectral resolution of $\sim 20,000$). Spectra were obtained in the $K+$ setting each year from 2014 to 2019 and in a special setting in the $K$-band ($K_{\mathrm{sp}}$) in 2021. The AO system on Subaru AO188 was employed in the observations. During the observations in 2014, 2016, and 2017, the LGS system was applied.

Table\,\ref{Tab:IRCS_Echelle} shows the echelle order numbers and wavelength coverage for the settings. Although seven echelle orders can be simultaneously observed, we utilized only three orders, namely, 25, 26 and 27, in the following analysis. This is mainly due to strong atmospheric absorption lines in other orders.

\begin{table}[htb]
\begin{center}
\caption{Wavelength coverage for echelle $K+$ and $K_{\mathrm{sp}}$ mode.
        \label{Tab:IRCS_Echelle}}
%\centering
\vspace{-0.3cm}
\begin{tabular}{cccc}
\hline
\hline
Setting & Order & Wavelength coverage\\
& & [$\mathrm{\AA}$] \\
\hline
$K+$ & 25 & $22500$--$23039$ \\
& 26 & $21634$--$22154$ \\
& 27 & $20833$--$21335$ \\ \hline
$K_{\mathrm{sp}}$ & 25 & $22324$--$22878$ \\
& 26 & $21466$--$21998$ \\
& 27 & $21670$--$21185$ \\
\hline
\end{tabular}
\end{center}
\end{table}

The details of the data reduction are described in Nishiyama et al.\cite{2018PASJ...70...74N}. The first process of data reduction involves flat fielding, bad pixel correction, and cosmic ray removal.
Flat field images were acquired using observations of a continuum source.

Prior to the extraction of spectra of S0-6, wavelength-calibrated stripe images,
where the $x$- and $y$-axes of the stripe images are wavelength and spatial position along the slit, respectively, were constructed.
Here, all OH emission lines along the slit direction are used for wavelength calibration of the stripe images.
The numbers of the OH lines used for the calibration are 6, 7, and 14 for the echelle orders 25, 26, and 27, respectively.

We extracted S0-6 spectra and then performed telluric correction of the S0-6 spectra using spectra of standard stars, which were bright early-A-type dwarfs.
Then, the S0-6 spectra were divided by the standard star spectra. We conducted continuum fitting for the telluric-corrected spectrum for each observation run. Finally, we combined all spectra to come up with a combined spectrum for each observed epoch.
To determine the metallicity, we combined all the RV-corrected spectra obtained in our observations. %%

We estimated the signal-to-noise (S/N) ratio for S0-6 employing the procedure described in Fukue et al.\cite{2015ApJ...812...64F}. Here, the S/N ratio is defined as the inverse of the noise for the normalized spectrum, $\sigma$, that is, S/N ratio $= \sigma^{-1}$.
Table\,\ref{Tab:spec_SNR} shows The S/N ratio for each observing run and each order.
The S/N ratios vary from 9 to 36 and depend not only on the number of the used data sets but also on the seeing condition of the observations and used guide star system, LGS or NGS.
The S/N ratios calculated for the combined spectra, 
composed of all the datasets from 2014 to 2021, were 57.3, 52.5, and 41.8, for echelle orders 27, 26, and 25, respectively. %%

\begin{table}[htb]
\begin{center}
\caption{Signal-to-noise ratio of spectra with $R = 20,000$. \label{Tab:spec_SNR}}
%\centering
\begin{tabular}{cccccc}
\hline
\hline
Date & & S/N ratio & \\ \hline
& order 27 & order 26 & order 25 \\
\hline
2014.379 & 23.1 & 22.9 & 21.1 \\
2015.635 & 22.5 & 23.5 & 20.4 \\
2016.381 & 18.9 & 18.2 & 16.4 \\
2017.341 & 22.1 & 20.7 & 19.6 \\
2017.344 & 33.1 & 36.2 & 30.6 \\
2017.347 & 30.1 & 30.2 & 28.9 \\
2017.603 & 11.2 & 13.3 & 9.3 \\
2017.609 & 23.1 & 23.1 & 14.1 \\
2018.087 & 22.1 & 24.5 & 19.6 \\
2021.420 & 28.1 & 12.3 & 22.7 \\
\hline
Combined$^{\mathrm{(a)}}$ & 57.3 & 52.5 & 41.8 \\
\hline
\end{tabular}
\end{center}
$^{\mathrm{(a)}}$ Combined spectra from 2014 to 2021.\\
\end{table}

%%%%%%%%%%%%%%%%%%%%%%%%%%%%%%%%%%%%%%%%%%%%%%%%%%%%%%%
%%%%%%%%%%%%%%%%%%%%%%%%%%%%%%%%%%%%%%%%%%%%%%%%%%%%%%%
\section{Radial velocities and location of S0-6} \label{sec:RV}

Absorption lines in its spectra were used in order to determine the RV for S0-6.
First, we roughly estimated the wavelength shift of the observed spectra using strong distinctive absorption lines in order 26. Second, the observed spectra were compared with a wavelength-shifted model spectrum of a late-type giant with a temperature of $\sim 4,000$\,K \cite{2019ApJ...872L..15H}.
If we found an absorption line that was also identified in the model spectrum,
we fitted the observed line with a Gaussian function
and calculated the RV of S0-6 from the wavelength difference
between the observed and model spectra.
Finally, the amount of RV correction was calculated using the IRAF {\it rvcorrect} task in order to obtain RVs in the Local Standard of Rest reference frame.

\begin{figure*}[h]
\begin{center}
    \includegraphics[width=10cm]{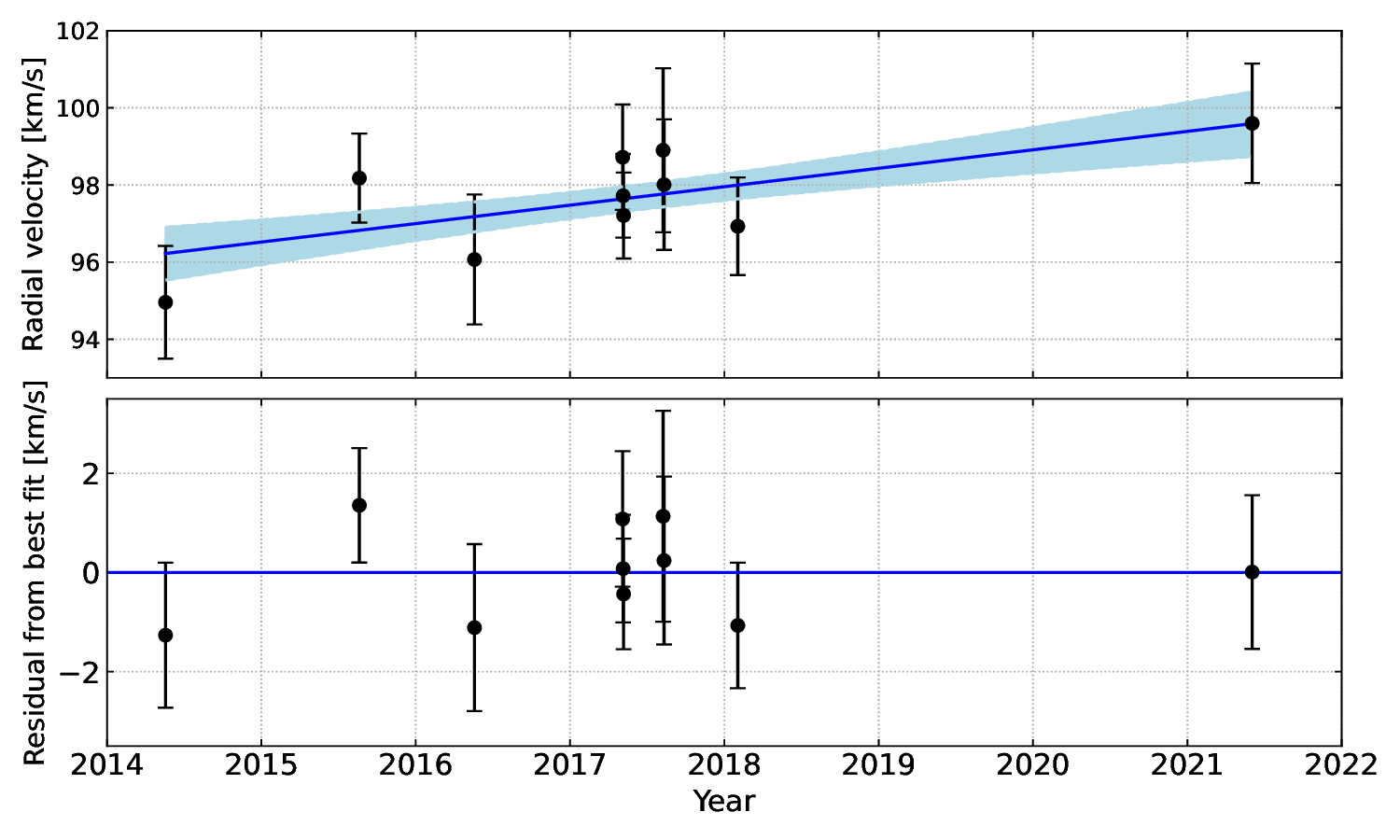}
\end{center}
    \caption{(Top) Radial velocity of S0-6 as function of time.
            The blue line represents the best-fit linear function.
            (Bottom) Residual from best-fit linear function
            shown in the top panel.
            \label{fig:RVplot}}
\end{figure*}

The time variation of the RV for S0-6 is shown in Fig.\,\ref{fig:RVplot}. The error bars include both statistical and systematic uncertainties (Table\,\ref{tab:RV_S0-6}). The statistical uncertainties are taken from the standard errors of the RV measurements using the absorption lines, where 21 to 45 lines are employed. The average statistical uncertainty is 1.25\,km\,s$^{-1}$. 

\begin{table*}[tb]
    \begin{center}
    \caption{Radial velocities and uncertainties for S0-6\label{tab:echelle}.}
    \begin{tabular}{ccccc} \hline
    Time & Number of lines & RV & Statistical uncertainty & Total uncertainty$^{(\mathrm{a})}$ \\
    (year) & & (km/s) & (km/s) & (km/s) \\ \hline
    2014.379 & 31 & 94.96 & 1.27 & 1.46 \\
    2015.635 & 31 & 98.18 & 0.90 & 1.15 \\ 
    2016.381 & 26 & 96.07 & 1.52 & 1.68 \\ 
    2017.341 & 31 & 98.72 & 1.16 & 1.37 \\ 
    2017.344 & 39 & 97.72 & 0.81 & 1.09 \\ 
    2017.347 & 44 & 97.21 & 0.85 & 1.12 \\ 
    2017.603 & 21 & 98.90 & 2.00 & 2.13 \\ 
    2017.609 & 28 & 98.01 & 1.53 & 1.69 \\ 
    2018.087 & 45 & 96.93 & 1.04 & 1.27 \\ 
    2021.420 & 34 & 99.60 & 1.37 & 1.55 \\ \hline
    \end{tabular}
    \label{tab:RV_S0-6}
%    \tablecomments{This }
%/Users/shogo/Dropbox/Documents/mypapers/S0-6/figs/RV/RVplot/S0-6_RVnoXerTotV2.dat
	\end{center}
	$^{(\mathrm{a})}$The systematic uncertainty of $0.72\,$km s$^{-1}$ is 
	quadratically summed to the statistical uncertainty.
\end{table*}

To understand the long-term stability of the RV measurements, we applied atmospheric OH emission lines. We extracted spectra at a position close to but different from S0-6 without subtracting background emission and fit the OH emission lines to measure wavelength-calibrated OH line wavelengths. By comparing them with the OH emission wavelengths measured in laboratories, we obtain a long-term stability of 0.72\,km\,s$^{-1}$ for the 7-year monitoring observations. We include this as a systematic uncertainty (Table\,\ref{tab:RV_S0-6}).
When the statistical and systematic uncertainties are summed quadratically, the average of the total uncertainties in our RV measurements is 1.45\,km\,s$^{-1}$. The small uncertainties shown above depict that this study is one of the most accurate and precise RV measurements that were conducted for stars in the central pc region of our galaxy.

Using a linear function, we fitted the RV plot, and the best-fit function is shown in Fig.\,\ref{fig:RVplot}.
The slope of the best-fit function is $0.48 \pm 0.20$\,km\,s$^{-1}$\,yr$^{-1}$.
This might suggest a marginal detection of the acceleration of S0-6 along the line of sight. 
The bottom panel in Fig.\,\ref{fig:RVplot} represents the residuals of the data points from the best-fit linear function. The standard deviation of the data points around the best-fit function is 0.92\,km/s, suggesting that the observed data points are well-fitted with a linear function.

The RV of S0-6 has been also measured for more than 13 years using the Keck telescope\cite{2023ApJ...948...94C}. Acceleration along the line of sight was also detected as $0.83 \pm 0.12$\,km s$^{-1}$ yr$^{-1}$, and the difference between the results is only $1.5 \sigma$, considering the uncertainties.
The positive value of the acceleration of S0-6 suggests that it is located in front of Sgr\,A* with its acceleration vector pointing away from us.

The marginal detection of acceleration allows us to estimate the three-dimensional 
distance $r$ of S0-6 from Sgr\,A*.
The equation of motion for S0-6 whose mass is $m_{*}$ is
\begin{equation}
    m_{*} \boldsymbol{a} = - G \frac{M m_{*}}{r^2} \frac{\boldsymbol{r}}{r},
\end{equation}
where $\boldsymbol{a}$ is the acceleration of S0-6, $M$ is the mass of Sgr\,A*, $\boldsymbol{r}$ is the position vector, and $G$ is the gravitational constant.
We measured the acceleration along the line of sight, $a_{z}$,
where $\boldsymbol{a} = (a_x, a_y, a_z)$
and $| a_{z} |$ is smaller than the 
three dimensional acceleration $| \boldsymbol{a} |$. 
Hence, we obtain the following:
\begin{equation}
    | a_{z} | < | \boldsymbol{a} | = G \frac{M}{r^2}.
\end{equation}
This enables us to calculate the upper limit of the distance $r$ using the following equation:
\begin{equation}
    r < \sqrt{\frac{GM}{| a_{z} |}}.
    \label{eq:dist}
\end{equation}
Substituting $G = 6.67 \times 10^{-11}$\,m$^3$kg$^{-1}$s$^{-2}$, 
$M = 4.23 \times 10^6 M_{\sun}$ \cite{2019PASJ...71..126S},
and $| a_{z} | = 0.48$\,km\,s$^{-1}$\,yr$^{-1}$ 
into equation [\ref{eq:dist}],
we acquire 
$r <0.2$\,pc $\approx 4 \times 10^4$\,AU.
This suggests that S0-6 is truly close to Sgr\,A*.

Note that on the basis of our current data set, the zero-acceleration hypothesis cannot be ruled out. The significance of the detection of $\boldsymbol{a}$ is still $2.4\,\sigma$. When fitting the observed RVs using a function without acceleration (i.e., slope $=0$), we obtain $\chi^2/\mathrm{d.o.f} = 7.53/9$.
This results in a p-value of 0.58, and we cannot rule out the zero-acceleration hypothesis for S0-6. Although the acceleration was detected using the Keck observations ($6.9\,\sigma$)\cite{2023ApJ...948...94C}, more observations are needed to conclusively confirm its detection using our own data.

%%%%%%%%%%%%%%%%%%%%%%%%%%%%%%%%%%%%%%%%%%%%%%%%%%%%%%%
%%%%%%%%%%%%%%%%%%%%%%%%%%%%%%%%%%%%%%%%%%%%%%%%%%%%%%%
\section{Stellar parameters of S0-6} \label{sec:metal}

%%%%%%%%%%%%%%%%%%%%%%%%%%%%%%%%%%%%%%%%%%%%%%%%%%%%%%%
\subsection{Data sets, {\it StarKit}, and calibration} \label{subsec:metal}

We constrained the stellar parameters of S0-6 by computing synthetic spectra and comparing them with the observed spectra. To compare the observed and synthetic spectra, we utilized the {\it StarKit} code developed by Kerzendorf \& Do (2015)\cite{2015zndo.....28016K}.
{\it StarKit} is a spectral fitting framework using Bayesian inference to determine the best-fit parameters.
{\it StarKit} simultaneously fits stellar parameters such as the following:
the effective temperature $T_{\mathrm{eff}}$, the surface gravity $\log g$, the overall metallicity of all elements [M/H], the $\alpha$ element abundance [$\alpha$/Fe], and the spectral continuum.

A synthetic spectral library is necessary for {\it StarKit} to determine
the set of stellar parameters.
Bentley et al.\cite{2022ApJ...925...77B} found that the BOSZ grid\cite{2017AJ....153..234B} provides
smaller offsets and scatter than other ones,
and we thus decided to utilize the BOSZ grid.
The parameter spaces covered by the BOSZ grid in our analysis are as follows:
$3500 \leq T_{\mathrm{eff}} [\mathrm{K}] \leq 35,000$,
$0.0 \leq \log g \leq +4.5$,
$-2.0 \leq \mathrm{[M/H]} \leq +0.75$, and
$-0.25 \leq [\alpha/\mathrm{Fe}] \leq +0.5$.

Upon fitting $K$-band stellar spectra with {\it StarKit}, we excluded the following absorption lines and wavelength ranges from the analysis. We did not use echelle order 25 for the analysis as order 25 includes CO absorption features at $\gtrsim 20900$\,\AA, and fitting the CO lines could introduce significant biases in the determinations of the parameters. The absorption lines due to Na, Ca, Sc, and V were known to be strong when compared with those for stars in the galactic disk or the solar neighborhood \cite{2007ApJ...669.1011C, 2018ApJ...855L...5D, 2018ApJ...866...52T}.
Thus, the lines associated with these four elements were excluded from the fit of the spectra.

%%%%%%%%%%%%%%%%%%%%%%%%%%%%%%%%%%%%%%%%%%%%%%%%%%%%%%%
%\subsection{Calibration using StarKit} \label{subsec:CalibStarkit}

\begin{table*}[tb]
\begin{center}
    \caption{Data for calibration sample\label{tab:calibstars}}
    \footnotesize
    \begin{tabular}{lccccc} \hline
Name & Spectral type & $T_{\rm{eff}}$ & $\log g$ & [Fe/H] & [$\alpha$/Fe] \\
 & & (K) & (dex) & (dex) & (dex) \\ \hline
BD$-01$\,2971 & M5 & $3587 \pm 19$ & $0.5^{(\mathrm{a})}$ & $-0.63 \pm 0.26$ & $0.023\pm0.222^{(\mathrm{b})}$ \\
2MASS J19213390+3750202   & K6-7 & $3734\pm97$ & $1.12\pm0.10$ & $+0.30\pm0.08$ & $0.068\pm0.012^{(\mathrm{c})}$ \\
%+3750202 \\
HD 787 & K4 & $3962 \pm 72$ & $1.12\pm0.31$ & $-0.04\pm0.13$ & $0.133\pm0.114^{(\mathrm{b})}$ \\
$\alpha$\,Boo (Arcturus, HD124897) & K1.5 & $4296\pm110$ & $1.66\pm0.29$ & $-0.53\pm0.11$ & $0.228\pm0.015^{(\mathrm{c})}$ \\
$\mu$\,Leo (HD\,85503) & K2 & $4494\pm93$ & $2.44\pm0.24$ & $+0.27\pm0.15$ & $0.019\pm0.009^{(\mathrm{c})}$ \\
$\alpha$\,Ari (HD\,12929) & K2 & $4587\pm65$ & $2.59\pm0.18$ & $-0.11\pm0.06$ & $0.06^{(\mathrm{b,d})}$ \\
$\delta$\,Dra (HD\,180711) & G9 & $4856\pm51$ & $2.69\pm0.27$ & $-0.12\pm0.10$ & $0.02\pm0.02^{(\mathrm{e})}$ \\
$\varepsilon$\,Leo (HD\,84441) & G1 & $5365\pm76$ & $2.08\pm0.29$ & $-0.05\pm0.20$ & -- \\ \hline
    \end{tabular}
    \end{center}
%    \tablecomments{
%   \tablenotetext{
    $^{(\mathrm{a})}$Only two measurements with the same results are found.
    $^{(\mathrm{b})}$[$\alpha$/Fe] $=$ ([Mg/Fe]$+$[Ca/Fe]$+$[Ti/Fe])/3.
    $^{(\mathrm{c})}$[$\alpha$/Fe] $=$ [$\alpha$/M]$+$[M/H]$-$[Fe/H], where [$\alpha$/M] and [M/H] are from APOGEE.
    $^{(\mathrm{d})}$[X/Fe] $=$ [X/H]$-$[Fe/H] $=$ $A$(X)$-\log(N_{\mathrm{X}}/N_{\mathrm{H}})_{\sun}-$[Fe/H],
    where $A$(X) is the abundance of element X
    and $N_{\mathrm{X}}$ and $N_{\mathrm{H}}$ are the number of atoms per unit volume
    for the element X and hydrogen, respectively.
    Here, $\log(N_{\mathrm{X}}/N_{\mathrm{H}})_{\sun}$ values for Mg, Ca, and Ti are
    taken from Asplund {et~al.}(2009)\cite{2009ARA&A..47..481A}.
    $^{(\mathrm{e})}$Taken from Tautvai{\v{s}}ien{\.{e}} {et~al.}(2020)\cite{2020ApJS..248...19T}.\\
\end{table*}

The stellar parameters determined by {\it StarKit} appear to depict a systematic offset in comparison to other studies \cite{2017MNRAS.464..194F, 2022ApJ...925...77B}.
We analyzed the stellar parameters for a calibration sample (Table \ref{tab:calibstars}) to evaluate the systematic offsets and the scatter when we use {\it StarKit} and IRCS $K$-band spectra.
The equivalent widths of Na, Ca, and CO absorption features in the $K$-band are known to be sensitive to $\log g$\cite{1997AJ....113.1411R}. However, the CO absorption features are only partly included in our observed spectrum.
Thus, we fixed $\log g$ for the calibration sample to values determined in past studies (Table \ref{tab:calibstars}).

Table\,\ref{tab:CalibSummary} shows the acquired parameter offsets between the {\it StarKit} and literature values. Here, we assume that the stellar [M/H] values are the same as the [Fe/H] values. These results suggest that the {\it StarKit} and IRCS spectra tend to underestimate $T_{\mathrm{eff}}$, [M/H], and [$\alpha$/Fe] compared to the literature values. These results are consistent with the studies by Bentley et al.\cite{2022ApJ...925...77B}, where they concluded that {\it StarKit} underestimates the three parameters.

\begin{table}[h]
\begin{center}
    \caption{
    Offsets between {\it StarKit} (ST) and literature (lit) values
    determined for different parameters of the calibration sample
    \label{tab:CalibSummary}}
    \begin{tabular}{cccc} \hline
Order &  $\Delta T_{\mathrm{eff}}^{(a)}$ &  $\Delta \mathrm{[M/H]}^{(b)}$ & $\Delta$[$\alpha$/Fe]$^{\mathrm{(c)}}$ \\ 
 & (K) & (dex) & (dex) \\ \hline
27 & $-199 \pm 196$ & $-0.30\pm0.15$ & $-0.13\pm0.27$ \\
26 & $-204 \pm 142$ & $-0.17\pm 0.06$ & $-0.18\pm0.06$ \\ \hline
        \end{tabular}
    \end{center}
%    \tablecomments{
    $^{\mathrm{(a)}}$ $\Delta T_{\mathrm{eff}} = T_{\mathrm{ST}} - T_{\mathrm{lit}}$
    $^{\mathrm{(b)}}$ $\Delta \mathrm{[M/H]} = \mathrm{[M/H]_{ST}} - {\mathrm{[Fe/H]_{lit}}}$
    $^{\mathrm{(c)}}$ $\Delta [\alpha/\mathrm{Fe}] = [\alpha/\mathrm{Fe}]_{\mathrm{ST}} - [\alpha/\mathrm{Fe}]_{\mathrm{lit}} $
%        }
        \vspace{-0.5cm}
\end{table}

%%%%%%%%%%%%%%%%%%%%%%%%%%%%%%%%%%%%%%%%%%%%%%%%%%%%%%%
\subsection{Analysis of S0-6 with {\it StarKit}} \label{subsec:metalS06}

We conducted spectral fitting of S0-6 using {\it StarKit}. The fit was performed for the spectra of order 27 and 26 separately. Since the wavelength range of our observed spectra is insensitive to $\log g$, we estimated $\log g$ by using Yonsei-Yale isochrones \cite{2004ApJS..155..667D}, in which we find correlations between $T_{\mathrm{eff}}$, the metallicity, and $\log g$. The late-type stars in the central 0\farcs5 region are likely to be old, 3--10\,Gyr
\cite{2019ApJ...872L..15H}; thus we used 5- and 10-Gyr isochrones, the difference of which is negligible.

S0-6 is a late-type giant, and its $T_{\mathrm{eff}}$ was observed to be $\sim 4,000$\,K \cite{2019ApJ...872L..15H}. Giant stars with $T_{\mathrm{eff}} \sim 4000$\, are expected to have a $\log g$ value in the range $\approx 0.8$ to $\approx 2.0$, if they are not very metal poor ($[\mathrm{Fe/H}] > -1$). \cite{2020MNRAS.494..396F} 
For the first attempt to identify the stellar parameters, we fit the spectra of S0-6 using {\it StarKit},
by constraining $\log g$ to $0.8 \leq \log g \leq 2.0$. We then obtained the first results of the fit for $T_{\mathrm{eff}}$, [M/H], and [$\alpha$/Fe]. On the basis of the first results and the systematic offset and uncertainties obtained in \S \ref{subsec:metal}, we applied a stronger constraint on $\log g$ using the Yonsei-Yale isochrones. With the stronger second constraint on $\log g$, we fit the spectra with {\it StarKit} and achieved a second set of results for $T_{\mathrm{eff}}$, [M/H], and [$\alpha$/Fe].
By repeating this procedure several times, we obtained stronger constraints on $\log g$ and more reliable stellar parameters. Finally, we fit the spectra with a fixed $\log g$.
The fixed values are $\log g = 0.9$ for order 27 and $1.2$ for order 26.

\begin{figure*}[tb]
\begin{center}
        \vspace{-0.5cm}
    \includegraphics[scale=0.45]{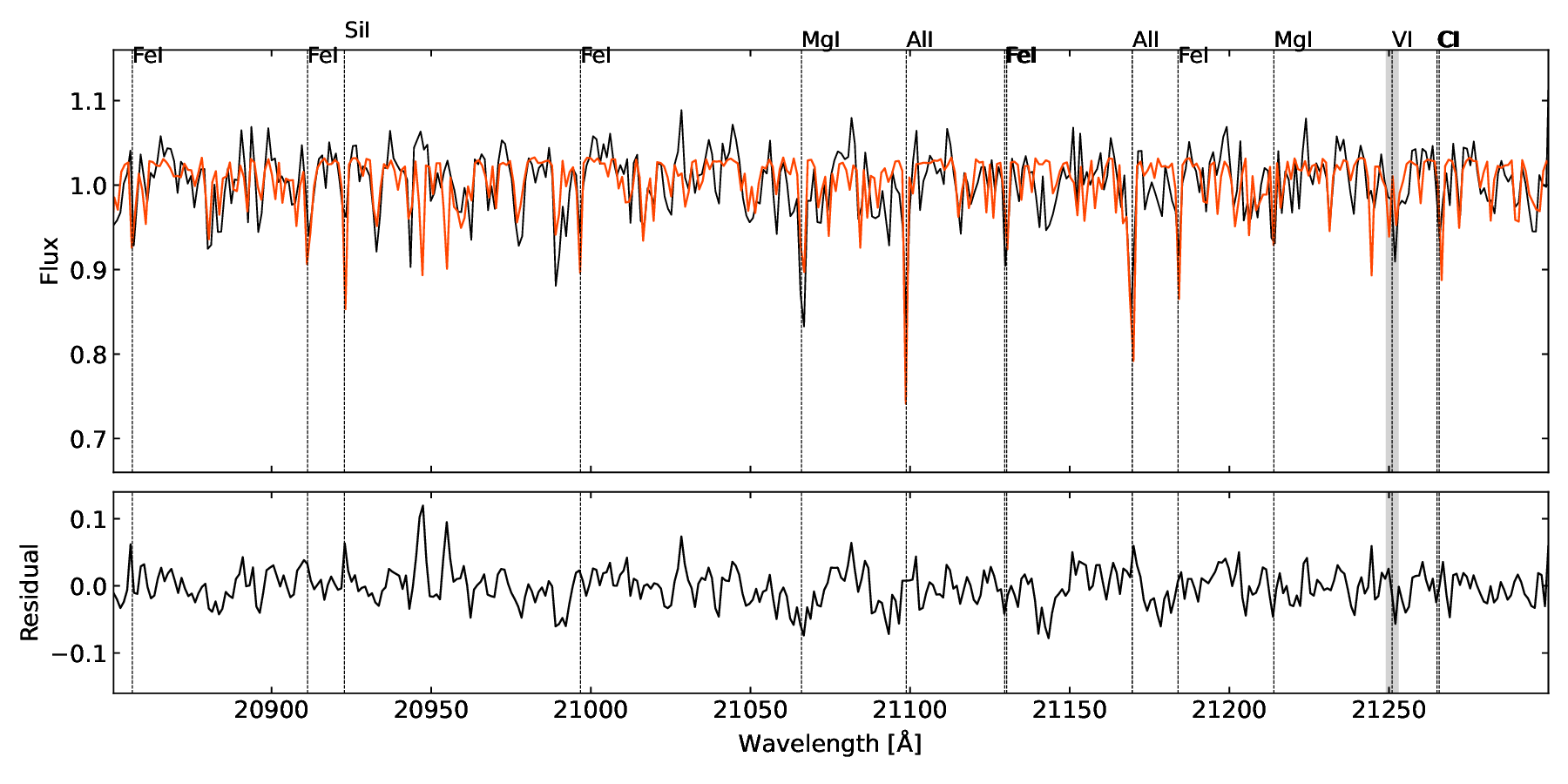}
    \includegraphics[scale=0.45]{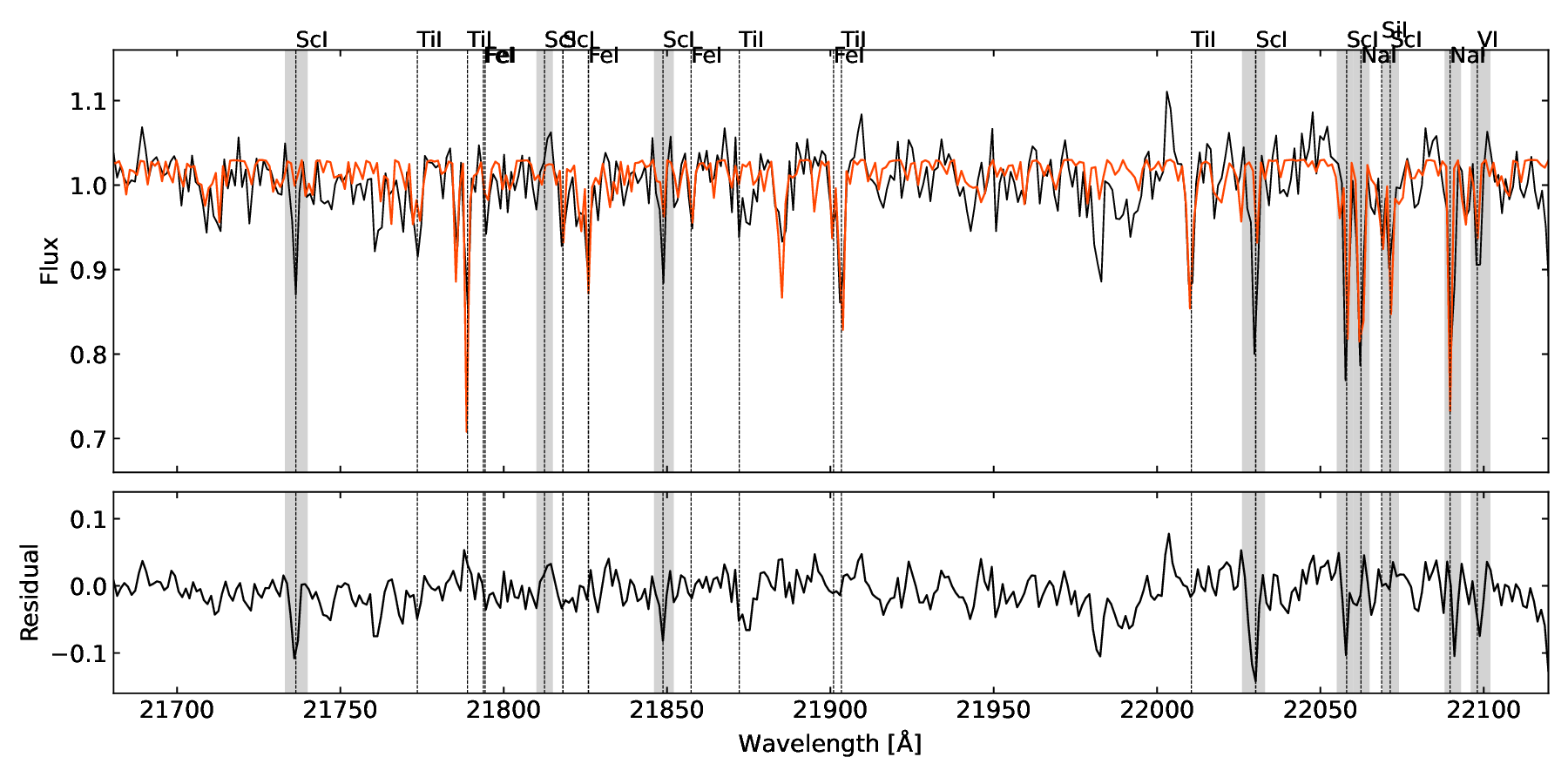}
    \caption{Observed spectra (black) and best-fitting model (red) of S0-6,
    	and residuals between observed and model spectra
	for order 27 (top) and 26 (bottom).
        The model parameters for order 27 
        are $T_{\mathrm{eff}} = 3630$\,K, 
        $\log g = 0.9$, [M/H]$ = -0.69$, and [$\alpha$/Fe] $= -0.23$,
        and those for order 26 are
        $T_{\mathrm{eff}} = 3790$\,K, 
        $\log g = 1.2$, [M/H]$ = -0.72$, and [$\alpha$/Fe] $= -0.22$.
        The BOSZ synthetic spectral library \cite{2017AJ....153..234B} 
        was applied in order to generate the model spectra.
        The vertical dashed lines represent known spectral lines.
        They are labeled on top of the panels.
        The grey regions of the spectra
        are excluded from the fitting process
        (see \S\,\ref{subsec:metal}).
        \label{fig:SpecFit}}
        \end{center}
\end{figure*}

Table\,\ref{Tab:result_specFit} lists the resulting stellar parameters.
Fig.\,\ref{fig:SpecFit} presents the spectra of S0-6 and the best-fitting models for orders 27 and 26.
The residuals between the observed and model spectra are also represented. 
The standard deviations of the residuals for orders 27 and 26 are 0.027 and 0.029, respectively. 
Considering the systematic offset shown in Table\,\ref{tab:CalibSummary}, the calibrated parameters are $T_{\mathrm{eff}} = 3749$\,K, $\mathrm{[M/H]} =-0.385$,  [$\alpha/\mathrm{Fe}]=-0.097$ (order 27), and $T_{\mathrm{eff}} = 3998$\,K, $\mathrm{[M/H]}=-0.546$, [$\alpha/\mathrm{Fe}]=-0.041$ (order 26).
Here, we assume that the systematic offsets are the same for good S/N spectra (the calibration samples) and for lower-quality spectra (S0-6).

\begin{table*}[htb]
\begin{center}
\caption{Results of spectral fitting
        \label{Tab:result_specFit}}
%\centering
\begin{tabular}{cccccccccc}
\hline
\hline
Order & \multicolumn{4}{c}{{\it StarKit} results} & & \multicolumn{4}{c}{After calibration} \\ \cline{2-5} \cline{7-10}
& $T_{\mathrm{eff}}$ & $\log g$ & [M/H] & [$\alpha$/Fe] & & $T_{\mathrm{eff}}$ & $\log g$ & [M/H] & [$\alpha$/Fe] \\
& (K) & fixed & (dex) & (dex) & & (K) & fixed & (dex) & (dex) \\ \hline
27 & $3630$ & $0.90$ & $-0.685$ & $-0.227$ & & $3749$ & $0.90$ & $-0.385$ & $-0.097$ \\ 
26 & $3794$ & $1.2$ & $-0.716$ & $-0.221$ & & $3998$ & $1.2$ & $-0.546$ & $-0.041$ \\ \hline
Average & & & & & & $3874$ & & $-0.466$ & $-0.069$ \\ \hline
\end{tabular}
\end{center}
\end{table*}

\begin{table*}[htb]
\begin{center}
\caption{Error budget in stellar parameter determination
        \label{Tab:uncertainty}}
%\centering
\begin{tabular}{ccccccccc}
\hline
\hline
%Aperture & 2 sets & calibration & standard deviation \\
%3 & \\
%4 & \\
& \multicolumn{2}{c}{$T_{\mathrm{eff}}$} & & \multicolumn{2}{c}{[M/H]} & & \multicolumn{2}{c}{[$\alpha$/Fe]} \\ \cline{2-3} \cline{5-6} \cline{8-9}
Order & 27 & 26 & & 27 & 26 & & 27 & 26 \\ \hline
Internal$^{\mathrm{(a)}}$ & $56$ & $163$ & & $0.20$ & $0.13$ & & $0$ & $0.019$ \\ % stderr
Calibration$^{\mathrm{(b)}}$ & $196$ & $142$ & & $0.15$ & $0.06$ & & $0.27$ & $0.06$ \\ \hline
Total in each order$^{\mathrm{(c)}}$ & $204$ & $216$ & & $0.25$ & $0.14$ & & $0.27$ & $0.063$ \\ \hline 
Average$^{\mathrm{(d)}}$ & \multicolumn{2}{c}{$210$} & & \multicolumn{2}{c}{$0.20$} & & \multicolumn{2}{c}{$0.196$} \\ %stderr
Standard error$^{\mathrm{(e)}}$ & \multicolumn{2}{c}{$124$} & & \multicolumn{2}{c}{$0.081$} & & \multicolumn{2}{c}{$0.028$} \\ \hline
Total uncertainty$^{\mathrm{(f)}}$ & \multicolumn{2}{c}{$244$} & & \multicolumn{2}{c}{$0.22$} & & \multicolumn{2}{c}{$0.20$} \\ %stderr
Final result & \multicolumn{2}{c}{$3870 \pm 240$} & & \multicolumn{2}{c}{$-0.47\pm0.22$} & & \multicolumn{2}{c}{$-0.07\pm0.20$} \\ \hline
\end{tabular}
\end{center}
    $^{\mathrm{(a)}}$Standard deviation of results for two spectra combined from two subsets.
    $^{\mathrm{(b)}}$Uncertainty in calibration, taken from Table\,\ref{tab:CalibSummary}.
    $^{\mathrm{(c)}}$Quadratic sum of internal and calibration uncertainties.
    $^{\mathrm{(d)}}$Error propagation in averaging results for orders 27 and 26.
    $^{\mathrm{(e)}}$Standard error of the results for orders 27 and 26, where the standard deviation of [M/H] for the two orders is divided by $\sqrt{2}$.
    $^{\mathrm{(f)}}$Quadratic sum of uncertainties in averaging and standard error.
\end{table*}

We summarize the uncertainties of our measurements in Table\,\ref{Tab:uncertainty}. We divided the observed data into two subsets to measure the internal uncertainty for each order.
The spectra in the subsets were combined, which resulted in two different spectra for each order. We performed fitting of the spectra with {\it StarKit} and determined the stellar parameters. The ``internal'' uncertainties in Table\,\ref{Tab:uncertainty} are the standard errors of the results of the two subsets. To estimate the total uncertainty in each order, the internal uncertainty and that derived in calibration were quadratically summed (``Total in each order'' in Table\,\ref{Tab:uncertainty}).
The ``Average'' uncertainties were determined by error propagation in averaging the results for the two orders. Besides the uncertainties described above, we include standard errors derived from the results of the two orders as an uncertainty. Finally, the uncertainties in ``average'' and ``standard error'' are quadratically summed in order to derive the total uncertainties. The derived stellar parameters of S0-6 are as follows:
$T_{\mathrm{eff}} = 3870 \pm 240$\,K,
[M/H] $= -0.47\pm0.22$, and
[$\alpha$/Fe] $= -0.07\pm0.20$.
For $T_{\mathrm{eff}}$, we find very good agreement between our result and the one of Habibi et al.\cite{2019ApJ...872L..15H}, $\approx 4000$--$4100$\,K.

%%%%%%%%%%%%%%%%%%%%%%%%%%%%%%%%%%%%%%%%%%%%%%%%%%%%%%%
\subsection{Analysis of S0-6 with Scandium lines and SME} \label{subsec:metalS06}

We know from earlier studies that the scandium energy level transition with the electron configuration 3d$^2$4s$-$3d4s4p, found near $2.2\,\mu$m, is sensitive to $T_\textrm{eff}$ of stars\cite{2018ApJ...866...52T,2020ApJ...894...26T}. This is because, toward 4000\,K, neutral scandium is becoming a minority species, which caused the scandium spectral features to be very sensitive to temperature. This is further strengthened by the fact that neutral scandium has 
a strong nuclear spin and hence features hyperfine structures, and for this energy level transition with a single s-electron at the lower level, the hyperfine structure is particularly strongly enhanced. Moreover, this energy-level transition is metastable, that is, has a long lifetime for spontaneous decay and therefore has a high electron population count. The combination of a hyperfine structure and high electron count makes the lines in the spectra particularly strong. This benefits establishing an empirical relation between the equivalent width of the scandium lines and $T_\textrm{eff}$ \cite{2020ApJ...894...26T}.

The empirical relation developed by Thorsbro et al.\cite{2020ApJ...894...26T} finds the equivalent width using the IRAF \textit{sbands} task with a width of 3 \AA\, around the scandium lines. The linear regression between the equivalent width and the temperatures results in a relationship with a standard deviation on the order of 50\,K.

Based on the scandium lines 21730\,\AA, 21812\,\AA, and 21842\,\AA, we employ the empirical relation between the equivalent width of the scandium lines and $T_\textrm{eff}$. 
Assuming that [Sc/Fe] is not likely to vary beyond 0.15\,dex \cite{2015A&A...577A...9B} and that the variation of 0.1\,dex in [Sc/Fe] leads to a variation of $\sim$100\,K \cite{2018ApJ...866...52T}, the uncertainty is $\sim$150\,K, which includes the statistical standard deviation from the empirical relation.
Hence, using the empirical relation, we found $T_\textrm{eff} = 3830 \pm 150$\,K. This agrees well with the results obtained using {\it StarKit}.

We also analyzed the star using {\it SME} \cite{1996A&AS..118..595V, 2012ascl.soft02013V} to determine chemical abundances. {\it SME} interpolates in a grid of one-dimensional MARCS atmosphere models \cite{2008A&A...486..951G}. These are hydrostatic model atmospheres in spherical geometry,  computed assuming LTE, chemical equilibrium, homogeneity, and conservation of the total flux.
The {\it SME} code has the advantage that it includes a flexible $\chi^2$ minimization tool to find the solution that best fits an observed spectrum in a pre-specified spectral window.

{\it SME} also has a powerful continuum normalization code, normalizing the spectra for every attempt to fit. In this way, {\it SME} can consider possible suppressed continuum levels that could exist due to the crowded spectra of cool stars. We found that for orders 26 and 27, we were able to do a good continuum normalization, but less so for order 25, thus confirming our previous choice not to use this order for the analysis. Order 25 has a CO-bandhead at the later part of the order, which extends beyond the edge of the order, causing it to have no anchor points for the continuum normalization on the higher wavelength range of the order.

To determine the chemical abundances, we need to have an accurate list of the atomic and molecular energy-level transitions. We use a list of atomic energy level transitions from a previous work by Thorsbro et al. \cite{2018IAUS..334..372T}, where wavelengths and line strengths (astrophysical $\log gf$-values) have been updated using the solar center intensity atlas \cite{1991aass.book.....L}.
We also include CN molecular lines, since the $K$-band is known for crowded CN-lines \cite{2014ApJS..214...26S}. Table~\ref{tab:lines} summarizes the lines used in this work.

\begin{table}[h]
\begin{center}
\caption{Lines used for abundance determinations.}
\label{tab:lines}
\begin{tabular}{ccc}
\hline
Wavelength in air (\AA) & $E_\mathrm{exc}$ (eV) & $\log gf$ \\
\hline\hline
\multicolumn{3}{l}{\rm Fe\,{\sc I}} \\
20948.086 & $-0.883$ & 6.1190 \\
20991.083 & $-3.019$ & 4.1427 \\
21123.885 & $-0.857$ & 6.0664 \\
21124.505 & $-1.647$ & 5.3342 \\
21756.929 & $-0.715$ & 6.2182 \\
21779.651 & $-4.298$ & 3.6398 \\
\hline
\multicolumn{3}{l}{\rm Mg\,{\sc I}} \\
21059.757 & $-0.709$ & 6.7791 \\
21061.095 & $-0.278$ & 6.7791 \\
\hline
\multicolumn{3}{l}{\rm Ca\,{\sc I}} \\
20937.903 & $-1.357$ & 4.6807 \\
20962.570 & $-0.740$ & 4.6814 \\
20972.529 & $-1.002$ & 4.6807 \\
\hline
\multicolumn{3}{l}{\rm Ti\,{\sc I}} \\
21767.610 & $-2.155$ & 2.5784 \\
21782.997 & $-1.170$ & 1.7489 \\
21897.437 & $-1.476$ & 1.7393 \\
22004.510 & $-1.950$ & 1.7335 \\ \hline
\end{tabular}
\end{center}
\end{table}

Table\,\ref{tab:ParS06SME} summarizes the results of the analysis with {\it SME}.
We found a metallicity value of [Fe/H] = $-0.4 \pm 0.15$. 
For the $\alpha$ element abundances, we found [Mg/Fe] = $-0.30 \pm 0.15$, 
[Ca/Fe] = $0.1 \pm 0.15$, and [Ti/Fe] = $-0.4 \pm 0.15$, 
giving an average [$\alpha$/Fe] = $-0.2 \pm 0.15$, 
based on an average of Mg, Ca, and Ti. 
These also agree well with the results from {\it StarKit}.

\begin{table*}[h]
\begin{center}
    \caption{
S0-6 parameters using the scandium method and SME
    \label{tab:ParS06SME}}
    \begin{tabular}{cccccc} \hline
 $T_{\mathrm{eff}}$ &  [Fe/H] & [Mg/Fe] & [Ca/Fe] & [Ti/Fe] & [$\alpha$/Fe]$^{\mathrm{(a)}}$ \\ 
(K) & (dex) & (dex) & (dex) & (dex) & (dex) \\ \hline
$3830 \pm 150$ & $-0.40\pm 0.15$ & $-0.30\pm0.15$ & $+0.10\pm0.15$ & $-0.40\pm0.15$ & $-0.20\pm0.15$\\ \hline
        \end{tabular}
    \end{center}
    $^{\mathrm{(a)}}$ [$\alpha$/Fe] = ([Ca/Fe] + [Mg/Fe] + [Ti/Fe])/3.
        \vspace{-0.5cm}
\end{table*}

%%%%%%%%%%%%%%%%%%%%%%%%%%%%%%%%%%%%%%%%%%%%%%%%%%%%%%%
%%%%%%%%%%%%%%%%%%%%%%%%%%%%%%%%%%%%%%%%%%%%%%%%%%%%%%%
\section{Discussion} \label{sec:discuccion}

%%%%%%%%%%%%%%%%%%%%%%%%%%%%%%%%%%%%%%%%%%%%%%%%%%%%%%%
%\subsection{A member of S-stars or NSC} \label{subsec:member}
\subsection{Location and age of S0-6} \label{subsec:LocAge}

We have marginally detected an acceleration in the RV for S0-6 (\S\,\ref{sec:RV}).
This suggests that S0-6 is accelerated by the SMBH
\cite{2023ApJ...948...94C}.
When we use the acceleration value and the mass of the SMBH, we obtain an upper limit for the distance of S0-6 from the SMBH of $\sim 0.2$\,pc $\approx 4 \times 10^4$\,AU. Hence, our result suggests that S0-6 is located close to the SMBH.
For the late-type stars within $\sim 1\arcsec$ from Sgr A*, $T_{\mathrm{eff}}$, bolometric magnitude, and spectral types are determined \cite{2019ApJ...872L..15H}, but no measurement of metallicity has been conducted. If S0-6 is indeed located at $0.2$\,pc from the SMBH, to our knowledge, this study is the first measurement of the chemical abundances of a late-type S-star.

No clear signal of binarity in observations of S0-6 was found. Fig.\,\ref{fig:RVplot} shows that the observed RVs are very well-fitted with a linear function, and no clear periodic signal is found in our observations. S0-6 was also observed with Keck/OSIRIS for 13\,years \cite{2023ApJ...948...94C}.
They find no significant periodic signal in the RVs of S0-6 and provide an upper limit of $0.1 M_{\sun}$ on the mass of a hypothetical companion object.

In order to estimate the age of S0-6, we plotted it on an HR diagram. The bolometric magnitude $M_{\mathrm{bol}}$ can be calculated to be $M_{\mathrm{bol}} = K - A_{K} - \mathrm{DM} + \mathrm{BC}_K$, where $K$, $A_{K}$, and DM are the observed $K$-band magnitude, the amount of the interstellar extinction in the $K$-band, and the distance modulus, respectively.
Here, we use $K = 13.95\pm0.04$
\cite{2019ApJ...871..103G},
$A_K = 2.46\pm0.03$
\cite{2010A&A...511A..18S},
and $\mathrm{DM} = 14.5\pm0.18$ 
\cite{2019PASJ...71..126S}.
We also used the equation
$\mathrm{BC}_{K} = 2.6 - (T_{\mathrm{eff}} - 3800)/1500$,\cite{2003ApJ...597..323B} to determine the bolometric correction $\mathrm{BC}_{K}$.
We finally obtained $M_{\mathrm{bol}} = -0.50\pm0.25$ and $-0.47\pm0.21$ for the {\it StarKit} and scandium method results, respectively.

\begin{figure*}[t]
\begin{center}
   \includegraphics[scale=0.70]{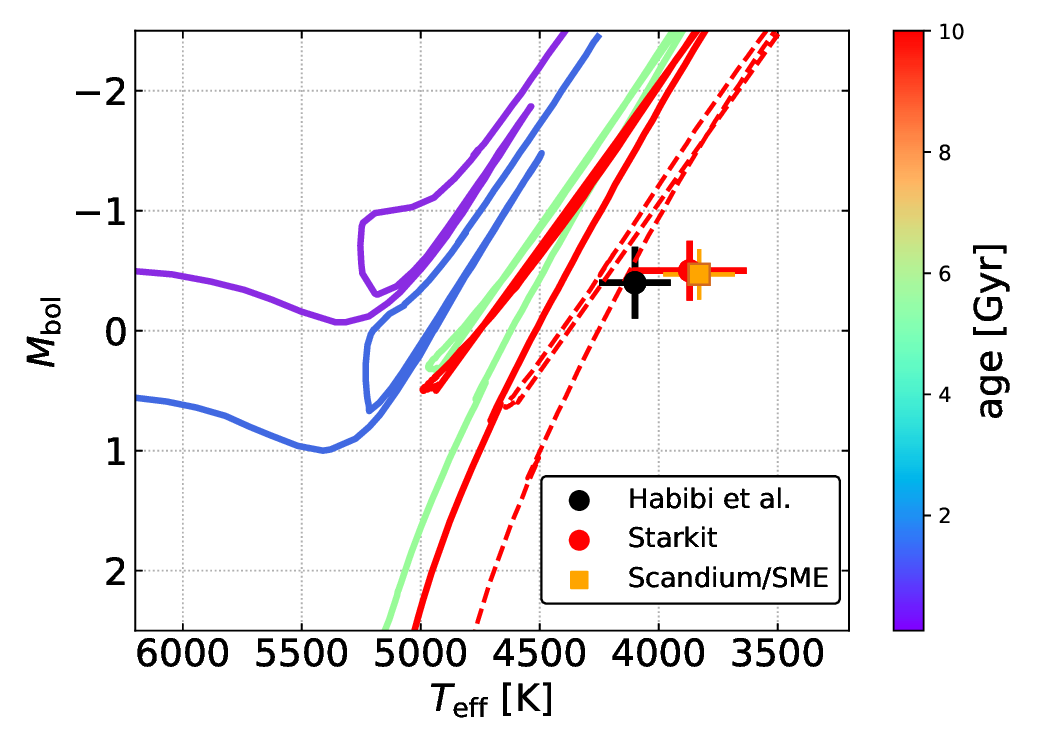}
    \caption{HR diagram showing 
	$T_{\mathrm{eff}}$ and $M_{\mathrm{bol}}$ of
    	S0-6 obtained by our study
        and Habibi et al. \cite{2019ApJ...872L..15H} (black circle).
 	The result of the {\it StarKit}, 
	and that of the scandium method and {\it SME} are represented
        by the red circle and orange square, respectively.
        Overlaid are $Z = 0.3Z_{\sun}$ theoretical isochrones \cite{2008A&A...482..883M}
        for 0.5, 1, 5, and 10\,Gyr (solid lines, from left to right).
        The dashed line is a theoretical 10\,Gyr isochrone for $Z = Z_{\sun}$.
    \label{fig:HRD}}
    \end{center}
\end{figure*}

S0-6 is plotted on an HR diagram in Fig.\,\ref{fig:HRD}.
The result of Habibi et al.\cite{2019ApJ...872L..15H} is also plotted, which agrees well with our results.
When S0-6 is compared with the theoretical $Z = 0.3Z_{\sun}$ isochrones\cite{2008A&A...482..883M} (solid lines in Fig.\,\ref{fig:HRD}),
it is located to the right of the 10\,Gyr isochrone, which suggests that S0-6 is an old star, as old as $\gtrsim 10$\,Gyr.

To further confirm the reliability of our $T_{\mathrm{eff}}$ determination, we estimated it using 
the line-depth-ratio (LDR) method. In the LDR method, ratios of the depths of two stellar absorption lines are employed to determine $T_{\mathrm{eff}}$.
Using well-studied bright stars such as $\alpha$\,Boo and $\mu$\,Leo, we found good LDR--$T_{\mathrm{eff}}$ relations for six line pairs in echelle order 26 (Nishiyama et al, in preparation for publication).
The line pairs are shown in Table\,\ref{tab:LDR}. The mean and standard deviation of $T_{\mathrm{eff}}$ using the six LDR-$T_{\mathrm{eff}}$ relations is $4100\pm90$\,K, where we show only the statistical uncertainty. This is slightly higher than $T_{\mathrm{eff}}$ derived by {\it StarKit} and the scandium method but consistent with them within the uncertainties.
Additionally, $T_{\mathrm{eff}}$ derived by {\it StarKit} for echelle order 26 is 3998\,K,
which agrees very well with that derived by the LDR method.

\begin{table}[h]
\begin{center}
    \caption{
    LDR pairs and derived $T_{\mathrm{eff}}$
    \label{tab:LDR}}
    \begin{tabular}{ccccc} \hline
Element &  Wavelength & Element & Wavelength & $T_{\mathrm{eff}}$$^{\mathrm{(a)}}$ \\ 
& (\AA) &  & (\AA) & (K) \\ \hline
Ti I & 22010.51 & Na I & 22089.69 & 4105 \\
Ti I & 22010.51 & Na I & 22062.42 & 4134 \\
Sc I & 22058.00 & Na I & 22089.69 & 3977 \\
Sc I & 22058.00 & Na I & 22062.42 & 4012 \\
Ti I & 21903.35 & Na I & 22089.69 & 4166 \\
Ti I & 21903.35 & Na I & 22062.42 & 4190 \\ \hline
        \end{tabular}
    \end{center}
    $^{\mathrm{(a)}}$ Mean and standard deviation of six $T_{\mathrm{eff}}$ is $4100\pm90$\,K.
\end{table}

Similar to S0-6, NSC stars distributed to the right and below the oldest isochrone were found 
\cite{2011ApJ...741..108P, 2016A&A...588A..49N, 2019ApJ...872L..15H}.
Chen et al. \cite{2023ApJ...944...79C} claimed that this discrepancy can be explained by assuming younger stars with higher metallicity. In this study, we found a similar discrepancy for an old and low-metal star, which suggests a previously unknown systematic uncertainty in observations and/or theoretical models.

%%%%%%%%%%%%%%%%%%%%%%%%%%%%%%%%%%%%%%%%%%%%%%%%%%%%%%%
\subsection{Chemical abundance of S0-6 and stars in our galaxy}

Recent observations \cite{2017MNRAS.464..194F,2020MNRAS.494..396F,2020ApJ...901L..28D}
found that there is a metal-poor stellar population ($\sim 7\,$\%) in the NSC,
whose mean metallicity is $[\mathrm{M/H}] \sim -0.5$.
Although Do et al. did not find stars with $[\mathrm{M/H}] \sim -0.5$
\cite{2015ApJ...809..143D},
Rich et al. found stars with $[\mathrm{Fe/H}] \sim -0.5$
\cite{2017AJ....154..239R}.
These suggest that stars with similar [M/H] values to S0-6 are not rare in the NSC,
and S0-6 may have the same origin as that of the NSC metal-poor population.
An interesting aspect of S0-6 is its closeness to Sgr\,A*,
because strong tidal force associated with the SMBH may provide an observable impact on
the nature of S0-6.

\begin{figure*}[t!]
\begin{center}
    \includegraphics[scale=0.6]{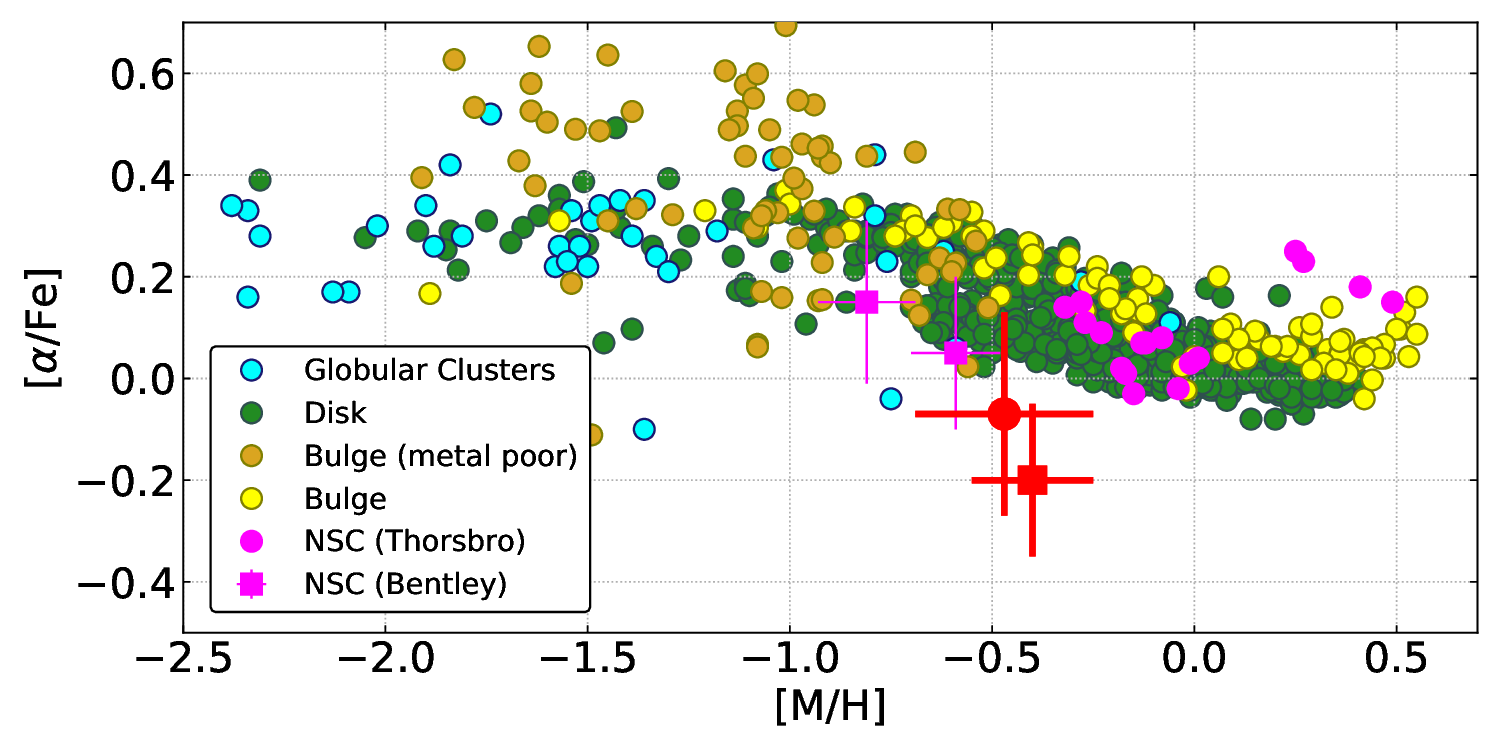}
    \includegraphics[scale=0.6]{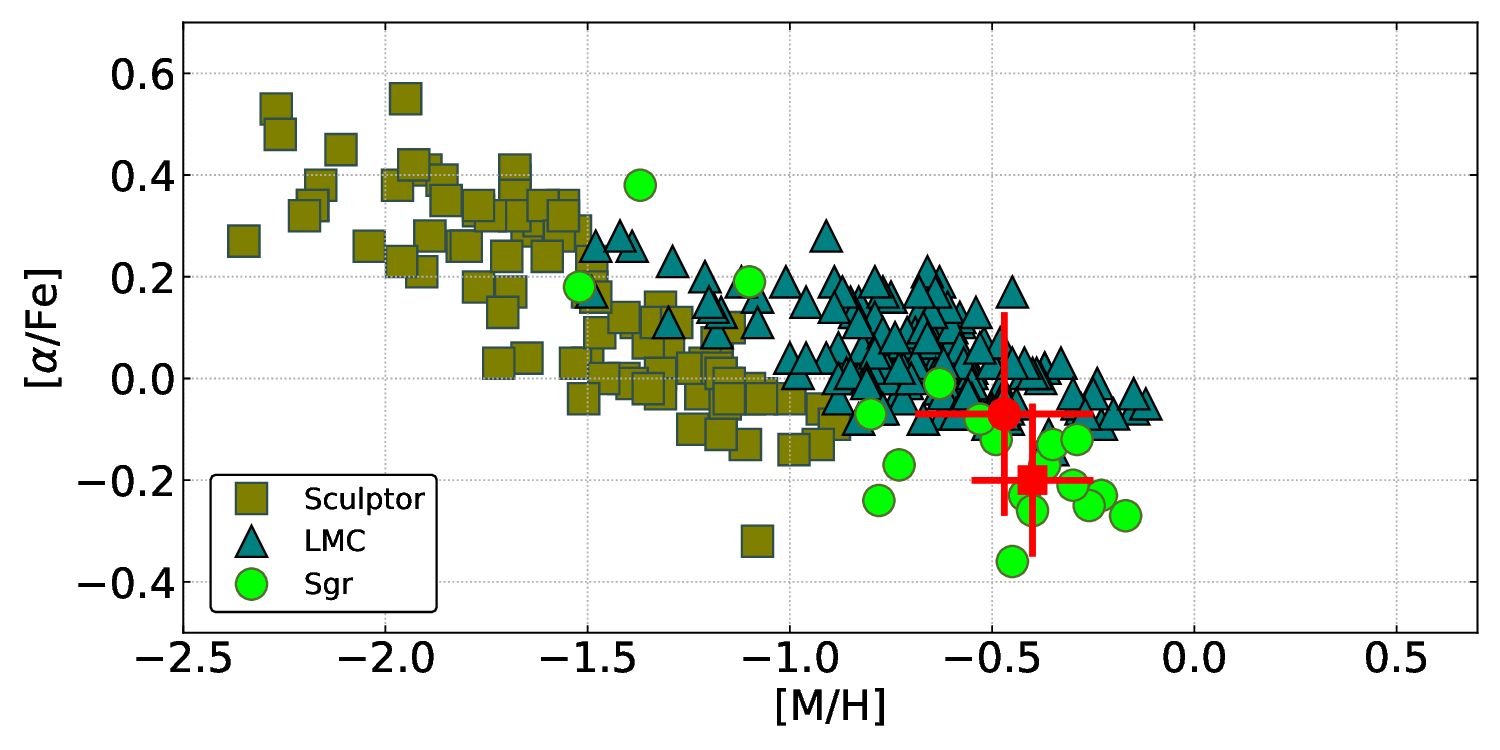}
    \caption{
    (Top) [$\alpha$/Fe] vs. [M/H] plot for stars and stellar systems in our galaxy.
    S0-6 is plotted as the red filled circle ({\it StarKit} result) and red square (SME result).
    The abundances of globular clusters are from Pritzl et al.\cite{2005AJ....130.2140P} 
    (cyan circles) assuming [Fe/H] $=$ [M/H].
    The abundances of disk stars, bulge stars, and bulge metal-poor stars are
    from Bensby et al.\cite{2014A&A...562A..71B} (green circles), 
    Bensby et al.\cite{2017A&A...605A..89B}(yellow circles),
    and Lucey et al.\cite{2022MNRAS.509..122L} (gold circles), respectively,
    assuming [Fe/H] $=$ [M/H]
    and [$\alpha$/Fe] $= ($[Mg/Fe]$+$[Si/Fe]$+$[Ca/Fe]$)/3$.
    The abundance of NSC stars are from Thorsbro et al.\cite{2020ApJ...894...26T} (magenta circles)
    assuming [Fe/H] $=$ [M/H] and [$\alpha$/Fe] $=$ [Si/Fe],
    and from Bentley et al.\cite{2022ApJ...925...77B} (magenta squares).
    (Bottom)
    [$\alpha$/Fe] vs. [M/H] plot for stars in nearby dwarf galaxies.
    S0-6 is plotted by the red-filled circle.
    The abundance of the Sculptor dwarf galaxy is 
    from Hill et al.\cite{2019A&A...626A..15H} (olive squares),
    and those for the Sgr dwarf galaxy are from Monaco et al.\cite{2005A&A...441..141M} (light green circles),
    assuming [Fe/H] $=$ [M/H] and [$\alpha$/Fe] = ([Mg/Fe] + [Ca/Fe])/2.
    The abundance of LMC is
    from Van der Swaelmen et al.\cite{2013A&A...560A..44V} (dark green triangle)
    assuming [Fe/H] $=$ [M/H] and [$\alpha$/Fe] = ([Mg/Fe] + [Si/Fe] + [Ca/Fe])/3.
%    Here we assume [Fe/H] $=$ [M/H].
        \label{fig:AFeFeHplot}}
    \end{center}
\end{figure*}

To examine the origin of S0-6 in more detail, we analyze its position in an [$\alpha$/Fe] vs. [M/H] diagram,
which is useful for clarifying the chemical evolution of stars and stellar systems. 
The top panel in Fig.\,\ref{fig:AFeFeHplot} plots S0-6 and stars in our galaxy.
It is clear that most of the stars in the galaxy with [M/H]$\lesssim 0.0$ show positive [$\alpha$/Fe].
We can observe two sequences in the diagram for disk stars (green circles) and bulge stars (yellow circles).
They overlap, but the bulge stars tend to have larger [$\alpha$/Fe] than the disk stars. At [Fe/H] $\sim -0.5$, the [$\alpha$/Fe] value for S0-6 is slightly smaller than that for the bulge stars, whereas it is consistent with the disk stars considering the observational uncertainty. There is a metal-poor population in the bulge, and for some elements, these metal-poor stars show different distributions in the diagrams from the other bulge stars and disk stars \cite{2022MNRAS.509..122L}.
Although the sequence is not clear in Fig.\,\ref{fig:AFeFeHplot}, S0-6 might be on the bulge metal-poor population sequence. We currently do not have enough samples to determine the distribution of late-type NSC stars in the diagram, because of the difficulty in observing such NSC stars. A pioneering study to determine the $\alpha$-element abundance of the late-type stars in the NSC region, a few pc from Sgr\,A*, was recently conducted by Thorsbro et al.\cite{2020ApJ...894...26T}, in which they determined [Si/Fe] for 15 stars (Fig.\,\ref{fig:AFeFeHplot}).
All of the stars measured by Thorsbro et al.\cite{2020ApJ...894...26T} show larger [M/H] and [$\alpha$/Fe] than S0-6.

The [M/H] and [$\alpha$/Fe] values were measured for two late-type stars within 1\,pc from Sgr\,A*
by Bentley et al.\cite{2022ApJ...925...77B}. They employed a method very similar to ours and found low-metal stars, whereas their [$\alpha$/Fe] values are positive.
One of the stars, NE1-1-003, shows a very similar chemical abundances to S0-6, [M/H]$=-0.59\pm0.11$ and [$\alpha$/Fe]$=0.05\pm0.15$.
Although the abundances of S0-6 are likely to be different from a large part of the stars in the NSC region, there might exist low [M/H] and low [$\alpha$/Fe] abundance stars in the region, for example, S0-6 and NE1-1-003.

%%%%%%%%%%%%%%%%%%%%%%%%%%%%%%%%%%%%%%%%%%%%%%%%%%%%%%%
\subsection{Chemical abundance of S0-6 and stars in dwarf galaxies}

Generally, stars in dwarf galaxies experience a different chemical evolution from those in our galaxy, 
and thus they show a different distribution in the [$\alpha$/Fe] vs. [M/H] diagram.
In the bottom panel in Fig.\,\ref{fig:AFeFeHplot}, 
stellar abundances for stars in three dwarf galaxies,
the Sagittarius dwarf spheroidal (Sgr) galaxy,
the Sculptor dwarf galaxy,
and the Large Magellanic Cloud (LMC)
are shown and compared with S0-6.

As can be seen, [M/H] of S0-6 is higher than that for 
the highest stellar population in the Sculptor dwarf galaxy. 
On the other hand, the position of S0-6 lies within the distributions of the stars of the LMC and the Sgr galaxy. 
Even so, we do not conclude that S0-6 was born in the LMC or the Sgr galaxy. 
The LMC is located at a distance of $\approx 50$\,kpc
and is orbiting around our galaxy.
A close encounter between the LMC and our galaxy likely occurred $\sim 1.5$\,Gyr ago, but the LMC has not been clearly tidally disrupted. The Sgr galaxy is experiencing strong and disruptive tidal interactions with our galaxy and has been almost totally disrupted, but has not yet been fully merged. Hence it is difficult to imagine that stars born in the LMC or Sgr galaxy would have migrated to the very center of our galaxy.

A more natural explanation for the origin might be that S0-6 was born in a dwarf galaxy and then migrated to the center of our galaxy. Recently, a minor, metal-poor population was found in the Galactic NSC
with kinematics distinct from the major, metal-rich population 
\cite{2017MNRAS.464..194F, 2020MNRAS.494..396F},
The metal-poor NSC population shows kinematics distinct from the metal-rich population
\cite{2017MNRAS.464..194F, 2020MNRAS.494..396F},
and its mass fraction was estimated to be $\sim7$--$18\,\%$
\cite{2020MNRAS.494..396F, 2017MNRAS.471.3617D}.
It might be natural to assume that S0-6 was born in the same cluster as the metal-poor population,
migrated to the center of our galaxy as a member of the cluster, and reached the central pc.
One of the scenarios to explain the low chemical abundances of S0-6 is that S0-6 was born as a star in an NSC of a dwarf galaxy, rather than a globular cluster of our galaxy.

%%%%%%%%%%%%%%%%%%%%%%%%%%%%%%%%%%%%%%%%%%%%%%%%%%%%%%%
%%%%%%%%%%%%%%%%%%%%%%%%%%%%%%%%%%%%%%%%%%%%%%%%%%%%%%%
\section{Conclusions} \label{sec:conclusion}

In this paper, we report the results of our NIR spectroscopic monitoring observations of the late-type S-star S0-6/S10 using the Subaru telescope and IRCS. Our main results are as follows.
We measured the RV of S0-6 with the average uncertainty of $1.45\,$km\,s$^{-1}$. This is one of the most precise RV measurements of a star in the central $1\arcsec$  region of our galaxy. We marginally detected an acceleration in the RV of S0-6 of $0.48 \pm 0.20$\,km\,s$^{-1}$\,yr$^{-1}$  from $2014$ to $2021$,
and obtained an upper limit for the distance of S0-6 from the SMBH of $0.2\,\mathrm{pc} \approx 4 \times 10^4\,$AU.
We determined the stellar parameters for S0-6 using the {\it StarKit} code
and the observed spectra.
The parameters are $T_{\mathrm{eff}} = 3870\pm240$\,K,
[M/H] $= -0.47\pm0.22$, [$\alpha$/Fe] $= -0.07\pm0.20$. 
We also determined stellar parameters using the scandium method and SME,
and obtained 
$T_{\mathrm{eff}} = 3830\pm150$\,K, [Fe/H]  $= -0.40\pm0.15$,
which are consistent with the {\it StarKit} results,
and both results suggest the very old age of S0-6, $\gtrsim 10$\,Gyr.
The abundance of other elements are:
[Ca/Fe]  $= +0.10\pm0.15$,  [Mg/Fe]$= -0.30\pm0.15$, and [Ti/Fe]  $= -0.40\pm0.15$. The chemical abundances and age suggest that S0-6 experienced a different chemical evolution from other stars in the center of our galaxy.

%%%%%%%%%%%%%%%%%%%%%%%%%%%%%%%%%%%%%%%%%%%%%%%%%%%%%%%
%%%%%%%%%%%%%%%%%%%%%%%%%%%%%%%%%%%%%%%%%%%%%%%%%%%%%%%
\section{Acknowledgements}

We wish to thank the Subaru Telescope staff for the support provided for our observations.
We thank Tuan Do, Rory Bentley, and Anja Feldmeier-Krause
for their kind assistance with the analysis using {\it Starkit}.
This work was supported by JSPS KAKENHI,
Grant in Aid for Challenging Exploratory Research 
(Grant Number 18K18760, 20H00178),
Grant-in-Aid for Scientific Research(A) 
(Grant Number 20H00178, 19H00695, 18H03720)
and Grant-in-Aid for Scientific Research(B) (50640843).
This work was supported by the Tohoku Initiative for Fostering GlobalResearchers for Interdisciplinary Sciences (TI-FRIS) of MEXT'sStrategic Professional Development Program for Young Researchers.
RS acknowledges financial support from the State Agency for Research of the Spanish MCIU through the ``Center of Excellence Severo Ochoa'' award for the Instituto de Astrof\'{i}sica de Andaluc\'{i}a (SEV-2017-0709) and from grant EUR2022-134031 funded by MCIN/AEI/10.13039/501100011033 and by the European Union NextGenerationEU/PRTR.
BT acknowledges the financial support from the Wenner-Gren Foundation (WGF2022-0041).This research is based on data collected at the Subaru Telescope,which is operated by the National Astronomical Observatory of Japan.We are honored and grateful for the opportunity to observe theUniverse from Maunakea, which has cultural, historical, and naturalsignificance in Hawaii.

%%%%%%%%%%%%%%%%%%%%%%%%%%%%%%%%%%%%%%%%%%%%%%%%%%%%%%%
%%%%%%%%%%%%%%%%%%%%%%%%%%%%%%%%%%%%%%%%%%%%%%%%%%%%%%%

%\profile[]{}

\end{document}